\documentclass[10pt,conference]{IEEEtran} 
\usepackage{amssymb}
\usepackage{amscd}
\usepackage{dsfont}
\usepackage{amsmath}
\usepackage{epsfig}
\usepackage{graphics}
\usepackage{psfrag}
\usepackage{rotating}
\usepackage{amsmath}
\usepackage{amsfonts}
\usepackage{url}
\usepackage{color}

\newtheorem{theorem}{Theorem}
\newtheorem{definition}[theorem]{Definition}
\newtheorem{proposition}[theorem]{Proposition}

\newcommand{\bs}{\boldsymbol}

\newcommand{\defn}{\stackrel{\triangle}{=} }
\newcommand{\ds}{\displaystyle}

\newcommand{\GameSE}{\mathcal{G}' = \left(\mathcal{K}, \left\lbrace \mathcal{S}_k \right\rbrace_{k \in \mathcal{K}},\left\lbrace u_{k}\right\rbrace_{ k \in\mathcal{K}}, \left\lbrace f_{k}\right\rbrace_{ k \in \mathcal{K}}\right)}
\newcommand{\GameNF}{\mathcal{G} = \left(\mathcal{K}, \left\lbrace\mathcal{S}_k \right\rbrace_{k \in \mathcal{K}},\left\lbrace u_{k}\right\rbrace_{ k \in \mathcal{K}}\right)}
\newcommand{\GameESE}{\mathcal{G}'' = \left(\mathcal{K}, \left\lbrace \mathcal{S}_k \right\rbrace_{k \in \mathcal{K}},\left\lbrace c_{k}\right\rbrace_{ k \in\mathcal{K}}, \left\lbrace f_{k}\right\rbrace_{ k \in \mathcal{K}}\right)}

\renewcommand{\baselinestretch}{.88}

\begin{document}
\title{Satisfaction Equilibrium: A General Framework for QoS Provisioning in Self-Configuring Networks}
\author{\IEEEauthorblockN{Samir M. Perlaza$^{1}$, Hamidou Tembine$^{2}$, Samson Lasaulce$^{2}$, M\'{e}rouane Debbah$^{3}$}
\IEEEauthorblockA{\\$^{1}$ France Telecom R\&D - Orange Labs Paris. France\\
Samir.MedinaPerlaza@orange-ftgroup.com\\
$^{2}$ Laboratoire des Signaux et Syst\`emes (LSS) - CNRS, SUPELEC, Univ. Paris Sud.  France\\
$\lbrace$Hamidou.Tembine, Samson.Lasaulce$\rbrace$@lss.supelec.fr\\
$^{3}$ Alcatel Lucent Chair in Flexible Radio - SUPELEC. France\\
Merouane.Debbah@supelec.fr}
}

\maketitle
\begin{abstract}
\boldmath
This paper is concerned with the concept of equilibrium and quality of service (QoS) provisioning in self-configuring wireless networks with non-cooperative radio devices (RD). In contrast with the Nash equilibrium (NE), where RDs are interested in selfishly maximizing its QoS, we present a concept of equilibrium, named satisfaction equilibrium (SE),  where RDs are interested only in guaranteing a minimum QoS. We provide the conditions for the existence and the uniqueness of the SE. Later, in order to provide an equilibrium selection framework for the SE, we introduce the concept of effort or cost of satisfaction, for instance, in terms of transmit  power levels, constellation sizes, etc. Using the idea of effort, the set of efficient SE (ESE) is defined. At the ESE, transmitters satisfy their minimum QoS incurring in the lowest effort. We prove that contrary to the (generalized) NE, at least one ESE always exists whenever the network is able to simultaneously support the individual QoS requests. Finally, we provide a fully decentralized algorithm to allow self-configuring networks to converge to one of the SE relying only on local information.
\end{abstract}

\section{Introduction}\label{SecIntroduction}

In the last decade, game theory has played a central role in the analysis of many problems regarding radio resource allocation and quality of service (QoS) provisioning in self-configuring wireless networks, see \cite{Lasaulce-Tutorial-09, Mackenzie-05} and references therein. These kind of problems can be modeled by non-cooperative games as long as radio devices (players) autonomously set up their transmission configuration (actions) to selfishly maximize their own QoS level (utility function). As a consequence, the concept of equilibrium introduced by Nash in \cite{Nash-1950} has been widely used. In the context of self-configuring networks, a Nash equilibrium (NE) is a network state at which radio devices cannot improve their QoS by unilaterally changing their transmission scheme. At the NE, each radio device attains the highest achievable QoS level given the transmission schemes of its counterparts.
\noindent
However, from a practical point of view, a radio device might be more interested in guaranteeing a minimum QoS rather than attaining the highest achievable one, due to several reasons. First, a reliable communication becomes possible only when certain parameters meet some specific conditions (minimum QoS requirement), e.g., minimum signal to  interference  plus noise ratio (SINR), minimum delay, etc. Second, higher QoS levels often imply higher efforts for the transmitter, e.g., higher transmit power levels, more complex signal processing, etc. Third, increasing the QoS for one communication often decreases the QoS of other communications. This reasoning implies that, in practical terms, the NE concept might fail to predict the effective network operating point and therefore its performance.
\noindent
In the presence of minimum QoS requirements, a more suited solution is the equilibrium concept introduced by Debreu in \cite{Debreu-1952} and nowadays known as generalized NE (GNE). In the context of self-configuring networks, a GNE is a state at which transmitters satisfy their QoS constraints and their performance cannot be improved by unilateral deviations (as in the NE).
Nonetheless, depending on the QoS metrics and network topology, the GNE might not exist \cite{Lasaulce-Tutorial-09}. In the case where it does, a transmitter always ends up achieving the highest achievable QoS, which is often costly, as mentioned above.
\noindent
In the most general case, one can consider that players aim to exclusively satisfy their constraints  instead of considering that players aim to maximize their own utility subject to a set of constraints. This reasoning leads to another type of equilibrium concept: any state of a given game where all players satisfy their own constraints is an equilibrium. Recently, Ross \emph{et al.} \cite{Ross-06} have formalized this concept for a particular type of constraints. Therein, such equilibrium is called satisfaction equilibrium (SE). In our scenario, an SE represents any network state where transmitters satisfy their QoS requirements, independently of their achieved QoS.

\noindent
In this paper, we generalize the idea of SE presented in \cite{Ross-06} such that it becomes independent of the type of constraints, and we present a brief discussion on its existence and uniqueness. Later, for each player, we arbitrarily define a function from its set of actions to the interval $\left[ 0,1\right]$. This function quantifies the effort of the player while using a given action. In this order of ideas, we introduce the concept of efficient SE (ESE). An ESE is a network state where all players satisfy their constraints by using the feasible action which requires the lowest effort. Assuming that the set of constraints is feasible, i.e., the minimum QoS requirements can be simultaneously supported by the network, we proved that contrary to the NE and GNE, at least one ESE always exists if the set of actions is finite, independently of the explicit form of the QoS metrics. Similarly, assuming the feasibility of the constraints, the existence of at least one ESE is also ensured when the set of actions is compact and the utility function is continuous over a linear space with finite dimension containing the set of SE action profiles. Finally, we present an algorithm which allows a set of transmitters to achieve an SE using only local information in a fully distributed fashion.

\section{Existing Game Theoretic Solutions for QoS Provisioning in Self-Configuring Networks}\label{SecGTModel}

As explained in Sec. \ref{SecIntroduction}, independently of the network topology (multiple access channels (MAC), interference channels (IC), etc.), the QoS provisioning problem in self-configuring network can be modeled by a static non-cooperative game. Consider a game in normal-form $\GameNF$. The set $\mathcal{K}$ represents the set of transmitters (players), and for all $k \in \mathcal{K}$, the set $\mathcal{S}_k$ represents the set of actions of transmitter $k$, e.g., a power allocation policy, a modulation scheme, etc. An action profile is a vector $\bs{s} = \left(s_{1}, \ldots, s_{K} \right) \in \mathcal{S}$,
where $\mathcal{S} = \mathcal{S}_1 \times \ldots \times \mathcal{S}_K$. We denote
by $\bs{s}_{-k} = \left(s_1, \ldots, s_{k-1},s_{k+1},\ldots, s_K\right)$ $\in \mathcal{S}_{-k} \defn \mathcal{S}_1 \times \ldots \times \mathcal{S}_{k-1} \times \mathcal{S}_{k+1} \times \ldots, \mathcal{S}_K$, the vector obtained by dropping off the $k$-th component of the vector $\bs{s}$. With a slight abuse of notation, we can write the vector $\bs{s}$ as $\left(s_k,\bs{s}_{-k}\right)$, in order to emphasize its $k$-th component.
For all $k \in \mathcal{K}$, the function $u_k: \mathcal{S} \rightarrow \mathds{R}$ is the utility function of transmitter $k$. This function determines how convenient (in the sense of the QoS) a given action $s_k \in \mathcal{S}_k$ is with respect to the actions adopted by all the other transmitters $\bs{s}_{-k}$. Hence, the higher the utility the better the action for a given transmitter. When the aim of each transmitter is to selfishly maximize its own utility function regardless of the utility obtained by its counterparts, a stable network configuration is the NE. An NE is defined as follows.
\begin{definition}[Pure Nash Equilibrium \cite{Nash-1950}]\label{DefNE} \emph{In the game $\GameNF$, an action profile $\boldsymbol{s} \in \mathcal{S}$ is a pure NE if it satisfies, for all $k \in \mathcal{K}$ and for all $\bs{s}'_k \in \mathcal{S}_k$,
\begin{equation}
u_k(s_k,\bs{s}_{-k}) \geqslant  u_k(s'_k,\bs{s}_{-k}).
\end{equation}}
\end{definition}
When constraints (QoS conditions) are imposed on the utilities that each transmitter obtains in the game $\mathcal{G}$, the NE is not longer a suited solution. In the presence of constraints, the set of actions each transmitter can take reduces to the set of actions which verifies the individual constraints given the actions adopted by the other transmitters. Let us characterize such a set of available actions by the correspondence $f_k: \mathcal{S}_{-k} \rightarrow 2^{\mathcal{S}_k}$ for each transmitter $k \in \mathcal{K}$ and denote the game with constraints by $\GameSE$. One of the solutions to the game $\mathcal{G}'$ is known as the generalized NE (GNE) \cite{Debreu-1952}, which is defined as follows:
\begin{definition}[Generalized NE \cite{Debreu-1952}]\label{DefGNE} \emph{An action profile $\boldsymbol{s}^* \in \mathcal{S}$ is a generalized Nash equilibrium (GNE) of the game $\GameSE$ if and only if
\begin{equation}\nonumber
\forall k \in \mathcal{K}, \quad s_k^* \in f_{k}\left(\bs{s}_{-k}^*\right) \mbox{ and }	
\end{equation}
\begin{equation}\nonumber
	\forall k \in \mathcal{K} \text{ and } \forall s_k \in f_k\left(\bs{s}_{-k}^*\right), \; u_k(s_k^*,\bs{s}_{-k}^*) \geqslant u_k(s_k,\bs{s}_{-k}^*).
\end{equation}
}
\end{definition}
Note that the classical definition of NE (Def. \ref{DefNE}) is obtained from Def. \ref{DefGNE}, when the action set of transmitter $k \in \mathcal{K}$  does not depend on the actions of the other transmitters, i.e., $\forall k \in \mathcal{K}$ and $\forall \bs{s}_{-k} \in \mathcal{S}_{-k}$, $f_k\left(\bs{s}_{-k}\right) = \mathcal{S}_k$, which means that no constraints are imposed on the utilities. In the next section, we introduce a new game solution which is also suited for the analysis of QoS provisioning in self-configuring networks.

\section{A New Game Solution: Satisfaction Equilibrium}\label{SecSE}

Consider now that the players in the game $\GameSE$ are exclusively interested on satisfying its own utility constraint, i.e., a given QoS condition. Here, the idea of satisfaction becomes intuitive: a player is said to be satisfied if it plays an action which satisfies its constraints. Once a player satisfies its individual constraints it has no interest in changing its action, and thus, an equilibrium is observed if all transmitters are simultaneously satisfied. We refer to this solution as satisfaction equilibrium and we define it as follows.
\begin{definition}[Satisfaction Equilibrium] \label{DefSE} \emph{An action profile $\bs{s}^+$ is a satisfaction equilibrium for the game $\GameSE$ if
\begin{equation}\label{EqSE}
\forall k \in \mathcal{K}, \quad s_k^+ \in f_{-k}\left(\bs{s}_{k}^+\right) .
\end{equation}
}
\end{definition}
Note that by taking the particular choice for all $k \in \mathcal{K}$, $f_k(\bs{s}_{-k}) = \left\lbrace s_k \in \mathcal{S}_k: u_k\left(s_k,\bs{s}_{-k}\right) \geqslant \Gamma_k \right\rbrace$,
where $\Gamma_k$ is the minimum utility level required by player $k$, then, Def. \ref{DefSE} coincides with the definition of SE provided in \cite{Ross-06}. However, in this paper we will refer to the SE concept as in Def. \ref{DefSE} for the sake of generality.
Let $\mathcal{S}_{\mathrm{SE}}$ be the set of SE of the game $\mathcal{G}'$. Hence,
\begin{equation}\label{EqFeasibleSet}
\mathcal{S}_{\mathrm{SE}} \defn \left\lbrace \bs{s} \in \mathcal{S}: \forall k \in \mathcal{K}, s_k \in f_k(\bs{s}_{-k}) \right\rbrace \subseteq \mathcal{S}.
\end{equation}
Let also $\mathcal{S}_{\mathrm{GNE}}$ be the set of GNE of the game $\mathcal{G}'$. Then, from Def. \ref{DefGNE} and Def. \ref{DefSE}, it follows  that
\begin{equation}
\mathcal{S}_{\mathrm{GNE}} \subseteq \mathcal{S}_{\mathrm{SE}} \subseteq  \mathcal{S},	
	\label{EqSetOfE}
\end{equation}
which verifies the intuition that the SE concept is less restrictive than the GNE concept.
\noindent
In the following, we analyze the existence, uniqueness and efficiency issues for the SE in the game $\mathcal{G}'$.

\subsection{Existence and Uniqueness of the Satisfaction Equilibrium}

The existence of an SE in the game $\GameSE$ mainly depends on the constraints imposed on the utility function, i.e.,  the set of correspondences $\lbrace f_{k} \rbrace_{k \in \mathcal{K}}$. For instance, let the correspondence $F: \mathcal{S} \rightarrow  2^{\mathcal{S}}$ be defined as follows: $F(\bs{s}) = \left(f_{1}\left(\bs{s}_{-k}\right),\ldots,f_{K}\left(\bs{s}_{-k}\right)\right)$. Then, an SE exists if and only if
\begin{equation}\label{EqExistence}
\exists \bs{s} \in \mathcal{S}:  \bs{s} \in \quad	F(\bs{s}) .
\end{equation}
This formulation allows us to use existing fixed point (FP) theorems to provide sufficient conditions for the existence of the SE. For instance, from Kakutani's FP theorem \cite{Kakutani-1941} we can write the following proposition.
\begin{proposition}[Existence of the SE]\label{PropExistence} \emph{In the game $\GameSE$, let the set of actions $\mathcal{S}$ be a non-empty, convex and compact set. Let also the correspondence $F(\bs{s})$ have a closed graph and be non-empty and convex in the set of action $\mathcal{S}$. Then, the game $\mathcal{G}'$ has at least one SE.
}
\end{proposition}

\noindent
Note that in Prop. \ref{PropExistence} no conditions (e.g., continuity) are imposed over the utility functions $\lbrace u_{k}\rbrace_{k \in \mathcal{K}}$, to ensure the existence of the SE. Indeed, from Def. \ref{DefSE}, it can be implied that a necessary and sufficient condition for the existence of an SE is the feasibility of the constraints, i.e., the existence of at least one action profile which simultaneously satisfies all the constraints. Note also that the feasibility condition is a necessary but not a sufficient condition for the existence of the GNE, which implies that one can observe games possessing at least one SE and no GNE. The converse is not true. This result implies that, by using the SE concept rather that the GNE concept, any achievable network performance can be fixed as the network operating point depending on the QoS requests (the set of functions $\lbrace f_{k} \rbrace_{k \in \mathcal{K}}$). Interestingly, this flexibility is not offered neither by the NE nor the GNE but by the SE.

\noindent
Finally, we underline the fact that the existence of one SE does not necessarily imply its uniqueness. Indeed, it is difficult to provide the conditions to observe a unique SE for a general set of correspondences $\lbrace f_{k} \rbrace_{k \in \mathcal{K}}$. However, as we shall see in Sec. \ref{SecSimulations}, the set of SE is often non-unitary and thus, an equilibrium selection process might be required. In the following section, we introduce a novel equilibrium selection for the case of SE.

\section{Equilibrium Selection: Efficient Satisfaction Equilibrium}

Assume now that the set of SE is non-empty and non-unitary. Hence, might a given SE be better than any other SE? To answer this question, consider that players care about the cost or effort of using a given action. For instance, using a higher transmit power level or using a more complex modulation scheme (in the sense of the size of the constellation) might require a higher energy consumption and thus, reduce the battery life time of the transmitters. Hence, high transmit power levels and complex modulations can be considered as costly actions. If players are able to measure their effort they incur when using a specific action, then it becomes natural to think that players would aim to be satisfied with the minimum effort. Following this reasoning the efficient SE (ESE) is defined as follows.

\begin{definition}[Efficient SE]\label{DefESE} \emph{Define a function $c_k: \mathcal{S}_k \rightarrow \left[0,1 \right]$ for all $k \in \mathcal{K}$ and consider the game $\GameSE$. For all $\left(k,s_k^*,s_k' \right) \in \mathcal{K}\times\mathcal{S}_k^2$, the action $s_k'$ is said to be more costly that action $s_k^*$ if  $c_{k}\left(s_k'\right)>c_{k}\left(s_k\right)$. An action profile $\bs{s}^*\in\mathcal{S}$ is an ESE if and only if
\begin{equation}\label{EqOPESE}
\forall k \in \mathcal{K}, \quad \bs{s}_k^{*} \in \ds\arg\min_{s_k \in f_k\left(\bs{s}_{-k}^{*}\right)} c_{k}\left(s_k\right).	
\end{equation}
Then, $\bs{s}^{*}$ is one of the efficient SE (ESE) of the game $\mathcal{G}'$.}
\end{definition}

From Def. \ref{DefESE}, it is implied that the set of ESE of the game $\GameSE$ coincides with the set of GNE of a non-cooperative game in normal-form denoted by $\mathcal{G}'' = \left\lbrace \mathcal{K}, \lbrace \mathcal{S}_k \rbrace_{k \in \mathcal{K}}, \lbrace c_k \rbrace_{\forall k \in \mathcal{K}},\lbrace f_k \rbrace_{\forall k \in \mathcal{K}}\right\rbrace$, where players aim to minimize their respective cost function $c_k$ subject to the set of constraints imposed over their utility functions $u_k$ and  represented by the function $f_k$.

An important remark on Def. \ref{DefESE} is that if all players assign the same cost to all their actions, then the set of ESE and SE are identical. This implies that the interest of an ESE is precisely that players can differentiate the costs of playing one action or another. Interestingly, the selection of the ESE is not based on the utilities obtained by the players but rather on their cost functions. This is because players are careless of their achieved utility as long as they are satisfied.

\subsection{Existence of an ESE}

Our first step to determine the existence of at least one ESE in the game $\mathcal{G}'$ is to show that the auxiliary game $\mathcal{G}''$ is an exact constrained potential game. We extend the definition of potential games in  \cite{Monderer-Shapley-1996} to exact constrained potential games as follows.
\begin{definition}[Exact Constrained Potential Game]\label{DefPG} \emph{Any game in normal form defined by the $4$-tuple $\left(\mathcal{K}, \left\lbrace\mathcal{S}_k\right\rbrace_{k \in \mathcal{K}}, \left\lbrace c_k\right\rbrace_{k \in \mathcal{K}},\left\lbrace f_k\right\rbrace_{k \in \mathcal{K}}\right)$ is an exact constrained potential game (PG) if there exists a function $\phi\left(\bs{s}\right)$ for all $\bs{s} \in \mathcal{S}_{\mathrm{SE}}$ such that for all players $k \in \mathcal{K}$ and for any pair of actions $\left( \bs{s}_k,\bs{s}'_k\right) \in \left\lbrace f_k\left(\bs{s}_{-k}\right) \right\rbrace^2$, it holds that
\begin{equation}\nonumber
 c_k(s_{k},\bs{s}_{-k}) - c_k(s'_{k},\bs{s}_{-k}) = \phi(s_{k},\bs{s}_{-k}) - \phi(s'_{k},\bs{s}_{-k}).
\end{equation}
}
\end{definition}
Before we continue, we clearly state that not all the properties of potential games \cite{Monderer-Shapley-1996} hold for the constrained potential games. As we shall see later, the best response dynamics might fail to converge to an equilibrium action profile.

Note that the effort function $c_k$ in the auxiliary game $\mathcal{G}''$ is arbitrary chosen by each player $k \in \mathcal{K}$ and it is independent of the actions taken by all the other players. Hence, following Def. \ref{DefPG}, it becomes clear that the game $\mathcal{G}''$ is an exact constrained potential game with potential function
\begin{equation}\label{EqPotential}
\phi\left( \bs{s} \right) = \ds\sum_{k = 1}^{K}c_{k}\left( s_k \right),
\end{equation}
if the set of SE (\ref{EqFeasibleSet}) is non-empty. This result leads us to the following proposition.
\begin{proposition}[Existence of the ESE]\label{PropExistenceESE} \emph{The game $\GameESE$, with $\mathcal{S}_k $ a finite set for all $k \in\mathcal{K}$, cost functions $c_k: \mathcal{S}_k \rightarrow \left[ 0,1\right]$ and a non-empty set $\mathcal{S}_{\mathrm{SE}}$, always has at least one ESE.}
\end{proposition}
The proof of Prop. \ref{PropExistenceESE} comes from the fact that by assumption, the domain of optimization in (\ref{EqOPESE}) is non-empty. Additionally, from Def. \ref{DefGNE} and Def. \ref{DefSE} it becomes clear that the set of solutions of the optimization problem in (\ref{EqOPESE}) is identical to the set of GNE of the game $\mathcal{G}''$ and, since $\mathcal{G}''$ is a potential game with finite sets of actions it always has at least one equilibrium in pure strategies (Lemma 2.3 in \cite{Monderer-Shapley-1996}).
Note that following the same argument, we can extend Prop. \ref{PropExistenceESE} for the case of compact and convex sets of actions. For instance, if for all $k \in \mathcal{K}$, $\mathcal{S}_k$ is compact and convex and the function $c_k$ is continuous over a finite dimensional linear space containing $\mathcal{S}_k$, then under these conditions, at least one ESE always exists.

\subsection{Uniqueness of the ESE}
As in the previous section, here we use the fact that the set of ESE of the game $\mathcal{G}'$ is identical to the set of GNE of the game $\mathcal{G}''$ with cost functions arbitrarily chosen by each player. Hence, since the game $\mathcal{G}''$ is an exact constrained potential game, we can state the following proposition.
\begin{proposition}[ESE in compact set of actions] \label{PropUniquenessESE1} \emph{The ESE of the game $\GameESE$ with cost functions $c_k: \mathcal{S}_k \rightarrow \left[0,1\right]$  is unique if the set $\mathcal{S}_{SE} \subseteq \mathcal{S}$ is non-empty, compact, and the (potential) function $\phi\left( \bs{s}\right) = \textstyle\sum_{k \in \mathcal{K}} c\left(s_k\right)$ is continuous and strictly convex over a linear space with finite dimensions containing $\mathcal{S}_{\mathrm{SE}}$.}
\end{proposition}

The proof of Prop. \ref{PropUniquenessESE1} follows from the fact that any minimum of the potential function $\phi$ in the set $\mathcal{S}_{\mathrm{SE}}$ is a GNE of the game $\mathcal{G}''$ (Def. \ref{DefGNE}). When the set $\mathcal{S}_{\mathrm{SE}}$ is compact, any GNE must be a potential minimizer, and since the potential is strictly convex (by assumption) the GNE is unique.

In the case of discrete sets of actions, one can relay on the Tarski's FP theorem \cite{Tarski-1955} to write the following proposition.
\begin{proposition}[ESE in finite discrete set of actions] \label{PropUniquenessESE2} \emph{Consider the potential game $\GameESE$ with cost potential function $\phi: \bs{s} \in \mathcal{S} \rightarrow \textstyle\sum_{k \in \mathcal{K}} c_k(s_k)$ and non-empty set $\mathcal{S}_{\mathrm{SE}}$. Assume that the correspondence $F\left(\bs{s}\right) = \left(f_1(\bs{s}_{-k}), \ldots, f_K(\bs{s}_{-k})\right)$ is monotone increasing in the sense that
\begin{eqnarray}
\nonumber
\forall \left(\bs{s},\bs{s}'\right) \in \mathcal{S}^2, \; \phi\left(\bs{s}\right) < \phi\left(\bs{s}'\right) \text{ implies } \\
\nonumber
\forall \left(\bs{s}^*,\bs{s}^+\right) \in F\left( \bs{s} \right) \times F\left( \bs{s}' \right), \; \phi\left(\bs{s}^*\right) < \phi\left(\bs{s}^+\right).
\end{eqnarray}
Then, the game $\mathcal{G}''$ has a unique ESE.
}
\end{proposition}
The proof of Prop. \ref{PropUniquenessESE2} stems from the fact that if the correspondence $F$ is monotone increasing in the sense discribed above, its set of fixed point solutions (which is non-empty, Prop. \ref{PropExistenceESE}) is a complete lattice. Thus, there is a unique minimizer of the potential function $\phi$.

\subsection{Determination of the ESE}\label{SecBRD}

To determine the set of ESE of the game $\GameSE$, one can simply solve the optimization problem in (\ref{EqOPESE}). However, this will require complete information for each player. For instance, a player might require the knowledge of the set of actions, actions actually being played, and parameters such as channel realizations and QoS requirements of all the other counterparts. Hence, this approach might not be practically appealing since transmitters only possess local information. In some particular scenarios, the ESE can be achieved in a fully decentralized fashion by using the best response dynamics (BRD)\cite{Lasaulce-Tutorial-09}, which in some cases requires a minimum feedback from the receivers. For instance, in interference channels where the utility function is the transmission rate, the set of actions is a compact set of power levels and the QoS constraint is a minimum transmission rate, the BRD might in some cases converge to an ESE \cite{Scutari-icassp-2006}\cite{PangScutari-IT-08}. However, in the presence of clipping actions, which we describe in the next section, then the BRD might not necessarily converge. In the next section, we present a general algorithm which is able to converge to any of the SE of the game $\mathcal{G}'$ but not necessarily to an ESE, requiring only the local information.

\section{Achieving Satisfaction Equilibria}\label{SecAlgorithms}
Now, we focus on the design of decentralized algorithms for allowing self-configuring wireless networks to achieve any SE (not necessarily an ESE) in the case when the QoS constraints can be written as  $f_k(\bs{s}_{-k}) = \left\lbrace s_k \in \mathcal{S}_k: u_k\left(s_k,\bs{s}_{-k}\right) \geqslant \Gamma_k \right\rbrace$, where $\Gamma_k$ is the minimum utility level required by player $k$. At most, we assume that a transmitter knows its own set of actions and is able to periodically observe its own achieved utility.

\noindent
Let us index the elements of each set $\mathcal{S}_k$, $\forall k \in \mathcal{K}$, with the index $n_k \in  \mathcal{N}_k \defn \lbrace 1, \ldots, |\mathcal{S}_k|\rbrace$, in any particular order. Denote by $s_k^{(n_k)}$ the $n_k$-th action of transmitter $k$. Assume that transmitter $k \in \mathcal{K}$ chooses their actions at instant $t>0$ following the discrete probability distribution $\bs{\pi}_k(t) = \left(\pi_{k,1}(t), \ldots,\pi_{k,\left|\mathcal{S}_k\right|}(t)\right)$, where $\pi_{k,n_k}(t)$ is the probability with which transmitter $k$ chooses its action $s_{k}^{(n_k)}$ at instant $t$.
Using this notation, we present the satisfaction equilibrium search algorithm (SESA), a slightly modified version of the algorithm presented in \cite{Chandramouli-08} (for the case of NE), which allows the convergence to an SE in a fully distributed fashion:\\
1) At time $t = 0$, all transmitters $k \in \mathcal{K}$ set up their initial action $s_{k}(0)$, following an arbitrary chosen probability distribution $\bs{\pi}_k(0)$.\\
2) At each time $t>0$, each transmitter $k \in \mathcal{K}$ computes $b_{k,t}=\frac{M_k + \hat{u}_{k,t-1}-\Gamma_k}{2 M_k}$, where $\hat{u}_{k,t}$ is the observed utility and $M_k$ is the highest utility transmitter $k$ can achieve (single user scenario). Then, it updates its actions as follows \begin{eqnarray}
	s_k(t) =\left\lbrace
					 \begin{array}{lcl} s_k(t-1) & \text{ if } & \hat{u}_{k,t}-\Gamma_k \geqslant 0 \\
																					 s_k(t) \sim \bs{\pi}_k(t) & \text{ otherwise. } & \end{array} 						  \right.\nonumber
\end{eqnarray}
and its probability distribution as follows, $\forall n_k \in \mathcal{N}_k,$
\begin{eqnarray}
	\bs{\pi}_{k,n_k}(t) =\left\lbrace
					 \begin{array}{lc} \bs{\pi}_{k,n_k}(t-1), &  \text{ if } \hat{u}_{k,t}-\Gamma_k \geqslant 0 \\
									   g_k(\bs{\pi}_k(t-1)) &  \text{ otherwise, } \end{array} 						  \right.\nonumber
\end{eqnarray}
Here,
$$g_k(\scriptstyle\bs{\pi}_{k,n_k}(t)\textstyle) = \scriptstyle \bs{\pi}_{k,n_k}(t) +  \lambda_{k,t} b_{k,t}\left(\mathds{1}_{\scriptscriptstyle\lbrace s_k(t) = s_{k}^{(n_k)}\rbrace} - \bs{\pi}_{k,n_k}(t) \right),$$
where $\forall k \in \mathcal{K}$, $\lambda_{k,t} = \frac{1}{t+1}$ is the learning rate of transmitter $k$.\\
3) If convergence is not achieved, then return to step (2).\\
It is important to remark that transmitters do not change their action dumbly. Conversely, at each action change, transmitters update their probability distribution so that higher probabilities are allocated to the actions which bring higher utilities and thus, reduces the time of convergence with respect to a time-invariant uniform probability distribution \cite{Ross-07}.
Before providing a result on the convergence of the SESA, we define  a clipping action as follows
\begin{definition}[Clipping Action]\label{DefClippingAction} \emph{In the game $\GameSE$, a player $k \in \mathcal{K}$ is said to have a clipping action $s_k$ if and only if
\begin{equation}
\forall \bs{s}_{-k} \in \mathcal{S}_{-k}, \quad s_k \in f_{k}\left(\bs{s}_{-k}\right).
\end{equation}}
\end{definition}
Once a player plays its clipping action, it remains indifferent to the actions of all the other players, since it is always satisfied. The existence of clipping actions in the game $\GameSE$ might inhibit the convergence of the SESA.
\begin{proposition} [Non-convergence of SESA]\label{PropNonConvergenceSESA} \emph{Assume the existence of at least one player with a clipping action in the game $\GameSE$ and denote it by $s_{k} \in \mathcal{S}_k$ for player $k$. Then, if there exists a player $j \in \mathcal{K}\setminus\lbrace k \rbrace$, for which $f_j\left(s_k,s_{-\lbrace j,k\rbrace}\right) = \emptyset$, $\forall \bs{s}_{-\lbrace j,k \rbrace} \in \mathcal{S}_{-\lbrace j,k \rbrace}$. Then, the SESA does not converge to an SE with strictly positive probability.}
\end{proposition}
The proof of Prop. \ref{PropNonConvergenceSESA} follows from the fact that at time $t$ before convergence, the probability of the clipping action $s_k$ is strictly positive and thus, player $k$ might play it. If so, by definition, there exist a player $j \neq k$ which would never be satisfied. Then, the SESA does not converge to any SE. On the contrary, if none of the players possesses a clipping action, the SESA converges to an SE  with probability one. This result comes from the fact that in the absence of clipping actions, there always exists a non-zero probability of visiting all possible action profiles. Once an SE action profile is visited, none of the players changes its action, and the convergence is observed.


\section{Case Study: Interference Channels}\label{SecSimulations}

In this section, we consider $K$ transmitter-receiver pairs simultaneously transmitting independent information and subject to mutual interference (interference channel). The QoS metric of each transmitter $k \in \lbrace 1 \ldots, K\rbrace$ is its transmission rate (utility function) and its set of actions is the set of different transmit power levels. Each transmitter aims to guarantee a minimum transmission rate denoted by  $\Gamma_k$ for player $k$. For all $(j,k) \in \lbrace 1, \ldots, K\rbrace^2$, let $h_{j,k}$ be the channel realization from transmitter $k$ to receiver $j$. At each channel use, channel coefficients are a realization of a complex circularly symmetric Gaussian random variable with zero mean and unit variance. We assume that channels are time-invariant during the whole transmission duration (e.g. a packet or frame duration). Let also $x_k(t)$ be the transmitted symbols of transmitter $k$ at time $t$. Here, $p_{k}(t) = \mathds{E}\left( x_k(t) x_k(t)^*\right) \leqslant	 p_{k,\max}$. The received signal at receiver $k$ can be written as $y_k(t) = h_{k,k} x_{k}(t) + \textstyle\sum_{j \neq k}^{K} h_{k,j} x_{j}(t) + w_k$, where $w_k$ is a random variable with variance $\sigma^2_k$ which represents the noise power at receiver $k$. The utility function of transmitter $k$ is
\begin{equation}
	u_{k}(t) = \log_{2}\left(1 + \frac{p_{k}(t)|h_{k,k}|^2}{\sigma^2_k + \textstyle\sum_{j \neq k}^K p_{j}(t)|h_{k,j}|^2} \right)\mbox{ [bps]},
\end{equation}
and the function $f_k$ is defined by $f_k(\bs{s}_{-k}) = \left\lbrace s_k \in \mathcal{S}_k: u_k\left(s_k,\bs{s}_{-k}\right) \geqslant \Gamma_k \right\rbrace$. The set of $N > 0$ (log-spaced) power levels of transmitter $k$ is
\begin{equation}
	 \mathcal{S}_k = \left\lbrace p_k =  p_{k,\max}\left(10^{-\frac{n}{N-1}\log_{10}(N)}\right), n \in \lbrace 0, \ldots, N-1\rbrace  \right\rbrace.
\end{equation}
For all $k \in \lbrace 1, \ldots, K \rbrace$, we define the effort function as the identity function, i.e., $c_k(q)= q$, for all $q \in \mathcal{S}_k$.

\begin{figure}
\centering
\includegraphics[width=\linewidth]{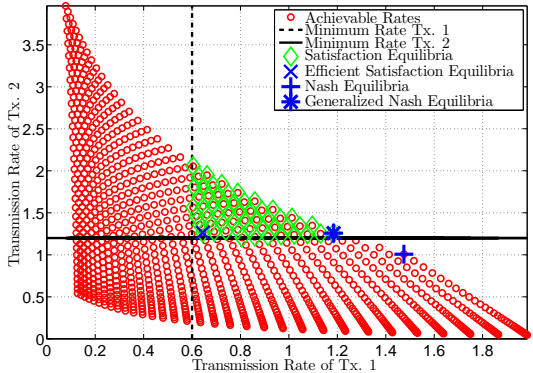}
\caption{Achievable rates and equilibria of the $2$-transmitter interference channel for a particular channel realization. Average SNR $\frac{p_{k,\max}}{\sigma^2_k} = 10$ dBs, $\bs{\Gamma} = \left(0.6, 1.2 \right)$ and $N = 32$. Channel realizations in Fig. \ref{FigConvergence} and Fig. \ref{FigUtilities} are the same.}
\label{FigUtilities}
\end{figure}

We model this scenario by the non-cooperative game in normal form $\GameSE$ for the case of $K = 2$ receiver-transmitter pairs. For all $k \in \lbrace 1, 2 \rbrace$, the average signal to noise ratio (SNR) is set to $\frac{p_{k,\max}}{\sigma^2_k} = 10$ dBs, $\Gamma = (0.6, 1.2)$ bps. In Fig. \ref{FigUtilities}, we plot the achievable transmission rates (achievable utilities) for both links and we identify the NE, GNE, SE and ESE. The (unique) NE is obtained when both transmitters use their maximum transmit power (strictly dominant action). The GNE is the solution where players maximize their utility and satisfy the QoS requirements. In this case, it is unique but might not be necessarily the case. The set of SE is the set $\mathcal{S}_{\mathrm{SE}}$. The ESE is unique (Prop. \ref{PropUniquenessESE1}) and corresponds to the solution to (\ref{EqOPESE}).
In Fig. \ref{FigUtilities} several statements are verified: (A) The unitary set of ESE is a subset of the set of SE, as suggested in (\ref{EqSetOfE}). (B) The GNE requires a higher transmit power than the efficient ESE, however, in both cases the transmitters satisfy the QoS constraints, and (C) the NE is not necessarily an SE.

\begin{figure}
\centering
\includegraphics[width=\linewidth]{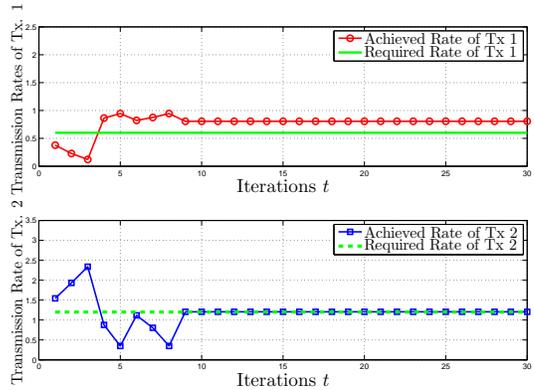}
\caption{Instantaneous achieved rates of transmitter $1$ (red) and $2$ (blue). Average SNR $\frac{p_{k,\max}}{\sigma^2_k} = 10$ dBs, $\bs{\Gamma} = \left(0.6, 1.2 \right)$ and $N = 32$. Channel realizations in Fig. \ref{FigConvergence} and Fig. \ref{FigUtilities} are the same.}
\label{FigConvergence}
\end{figure}

\noindent
In Fig. \ref{FigConvergence}, we plot the achieved transmission rate of both links at each instant $t$ when SESA is used. Therein, it becomes clear that even though a transmitter is satisfied, and thus does not change its action, its achieved rate changes due to the actions of the other transmitters. Once both transmitters are satisfied, then none of them changes its transmit powers.
%
%

\section{Conclusions}\label{SecConclusions}
We presented a new framework for QoS provisioning in self-configuring networks based on the concept of SE and ESE, both inspired from the game theory domain. The practical pertinence of these concepts is clearly evidenced here. However, several problems remain to be solved. In the one hand, a general algorithm for converging to an ESE is still unknown. In the other hand, as long as the network can satisfy the QoS requirements, our approach provides a solution. However, in the converse case, an approach on mixed strategies can be used to satisfy at least in expectation the QoS requirements. We let these two issues as interesting tracks for further works on applying SE and ESE in self-configuring networks.
\bibliographystyle{IEEEtran}
\bibliography{DSA-Library,GT}

\end{document}